\documentclass[aps,prd,reprint,superscriptaddress,showpacs,showkeys]{revtex4-1}

\usepackage{graphicx}
\usepackage{natbib}
\usepackage{color}

\usepackage{enumitem}
\setenumerate{itemindent=2em,leftmargin=0pt,itemsep=0pt,partopsep=0pt}
\setitemize[1]{itemindent=2em,leftmargin=0pt,itemsep=2pt,partopsep=0pt,topsep=0pt}

\begin{document}


\title{Limits on light WIMPs with a 1 kg-scale germanium detector at 160 eVee physics threshold at the China Jinping Underground Laboratory}


\affiliation{Key Laboratory of Particle and Radiation Imaging (Ministry of Education) and Department of Engineering Physics, Tsinghua University, Beijing 100084}
\affiliation{Department of Physics, Tsinghua University, Beijing 100084}
\affiliation{Institute of Physics, Academia Sinica, Taipei 11529}
\affiliation{NUCTECH Company, Beijing 100084}
\affiliation{YaLong River Hydropower Development Company, Chengdu 610051}
\affiliation{College of Physical Science and Technology, Sichuan University, Chengdu 610064}
\affiliation{Department of Nuclear Physics, China Institute of Atomic Energy, Beijing 102413}
\affiliation{School of Physics, Nankai University, Tianjin 300071}
\affiliation{Department of Physics, Banaras Hindu University, Varanasi 221005}

\author{L.T.~Yang}
\affiliation{Key Laboratory of Particle and Radiation Imaging (Ministry of Education) and Department of Engineering Physics, Tsinghua University, Beijing 100084}
\affiliation{Department of Physics, Tsinghua University, Beijing 100084}
\author{H.B.~Li}
\altaffiliation{Participating as a member of TEXONO Collaboration}
\affiliation{Institute of Physics, Academia Sinica, Taipei 11529}
\author{Q.~Yue}
\email{Corresponding author: yueq@mail.tsinghua.edu.cn}
\affiliation{Key Laboratory of Particle and Radiation Imaging (Ministry of Education) and Department of Engineering Physics, Tsinghua University, Beijing 100084}
\author{K.J.~Kang}
\affiliation{Key Laboratory of Particle and Radiation Imaging (Ministry of Education) and Department of Engineering Physics, Tsinghua University, Beijing 100084}
\author{J.P.~Cheng}
\affiliation{Key Laboratory of Particle and Radiation Imaging (Ministry of Education) and Department of Engineering Physics, Tsinghua University, Beijing 100084}
\author{Y.J.~Li}
\affiliation{Key Laboratory of Particle and Radiation Imaging (Ministry of Education) and Department of Engineering Physics, Tsinghua University, Beijing 100084}
\author{H.T.~Wong}
\altaffiliation{Participating as a member of TEXONO Collaboration}
\affiliation{Institute of Physics, Academia Sinica, Taipei 11529}
\author{M.~A\v{g}artio\v{g}lu}
\altaffiliation{Participating as a member of TEXONO Collaboration}
\affiliation{Institute of Physics, Academia Sinica, Taipei 11529}
\author{H.P.~An}
\affiliation{Key Laboratory of Particle and Radiation Imaging (Ministry of Education) and Department of Engineering Physics, Tsinghua University, Beijing 100084}
\affiliation{Department of Physics, Tsinghua University, Beijing 100084}
\author{J.P.~Chang}
\affiliation{NUCTECH Company, Beijing 100084}
\author{J.H.~Chen}
\altaffiliation{Participating as a member of TEXONO Collaboration}
\affiliation{Institute of Physics, Academia Sinica, Taipei 11529}
\author{Y.H.~Chen}
\affiliation{YaLong River Hydropower Development Company, Chengdu 610051}
\author{Z.~Deng}
\affiliation{Key Laboratory of Particle and Radiation Imaging (Ministry of Education) and Department of Engineering Physics, Tsinghua University, Beijing 100084}
\author{Q.~Du}
\affiliation{College of Physical Science and Technology, Sichuan University, Chengdu 610064}
\author{H.~Gong}
\affiliation{Key Laboratory of Particle and Radiation Imaging (Ministry of Education) and Department of Engineering Physics, Tsinghua University, Beijing 100084}
\author{L.~He}
\affiliation{NUCTECH Company, Beijing 100084}
\author{J.W.~Hu}
\affiliation{Key Laboratory of Particle and Radiation Imaging (Ministry of Education) and Department of Engineering Physics, Tsinghua University, Beijing 100084}
\author{Q.D.~Hu}
\affiliation{Key Laboratory of Particle and Radiation Imaging (Ministry of Education) and Department of Engineering Physics, Tsinghua University, Beijing 100084}
\author{H.X.~Huang}
\affiliation{Department of Nuclear Physics, China Institute of Atomic Energy, Beijing 102413}
\author{L.P.~Jia}
\affiliation{Key Laboratory of Particle and Radiation Imaging (Ministry of Education) and Department of Engineering Physics, Tsinghua University, Beijing 100084}
\author{H.~Jiang}
\affiliation{Key Laboratory of Particle and Radiation Imaging (Ministry of Education) and Department of Engineering Physics, Tsinghua University, Beijing 100084}
\author{H.~Li}
\affiliation{NUCTECH Company, Beijing 100084}
\author{J.M.~Li}
\affiliation{Key Laboratory of Particle and Radiation Imaging (Ministry of Education) and Department of Engineering Physics, Tsinghua University, Beijing 100084}
\author{J.~Li}
\affiliation{Key Laboratory of Particle and Radiation Imaging (Ministry of Education) and Department of Engineering Physics, Tsinghua University, Beijing 100084}
\author{X.~Li}
\affiliation{Department of Nuclear Physics, China Institute of Atomic Energy, Beijing 102413}
\author{X.Q.~Li}
\affiliation{School of Physics, Nankai University, Tianjin 300071}
\author{Y.L.~Li}
\affiliation{Key Laboratory of Particle and Radiation Imaging (Ministry of Education) and Department of Engineering Physics, Tsinghua University, Beijing 100084}
\author{F.K.~Lin}
\altaffiliation{Participating as a member of TEXONO Collaboration}
\affiliation{Institute of Physics, Academia Sinica, Taipei 11529}
\author{S.T.~Lin}
\affiliation{College of Physical Science and Technology, Sichuan University, Chengdu 610064}
\author{S.K.~Liu}
\affiliation{College of Physical Science and Technology, Sichuan University, Chengdu 610064}
\author{Z.Z.~Liu}
\affiliation{Key Laboratory of Particle and Radiation Imaging (Ministry of Education) and Department of Engineering Physics, Tsinghua University, Beijing 100084}
\author{H.~Ma}
\email{Corresponding author: mahao@mail.tsinghua.edu.cn}
\affiliation{Key Laboratory of Particle and Radiation Imaging (Ministry of Education) and Department of Engineering Physics, Tsinghua University, Beijing 100084}
\author{J.L.~Ma}
\affiliation{Key Laboratory of Particle and Radiation Imaging (Ministry of Education) and Department of Engineering Physics, Tsinghua University, Beijing 100084}
\author{H.~Pan}
\affiliation{NUCTECH Company, Beijing 100084}
\author{J.~Ren}
\affiliation{Department of Nuclear Physics, China Institute of Atomic Energy, Beijing 102413}
\author{X.C.~Ruan}
\affiliation{Department of Nuclear Physics, China Institute of Atomic Energy, Beijing 102413}
\author{B.~Sevda}
\altaffiliation{Participating as a member of TEXONO Collaboration}
\affiliation{Institute of Physics, Academia Sinica, Taipei 11529}
\author{V.~Sharma}
\altaffiliation{Participating as a member of TEXONO Collaboration}
\affiliation{Institute of Physics, Academia Sinica, Taipei 11529}
\affiliation{Department of Physics, Banaras Hindu University, Varanasi 221005}
\author{M.B.~Shen}
\affiliation{YaLong River Hydropower Development Company, Chengdu 610051}
\author{L.~Singh}
\altaffiliation{Participating as a member of TEXONO Collaboration}
\affiliation{Institute of Physics, Academia Sinica, Taipei 11529}
\affiliation{Department of Physics, Banaras Hindu University, Varanasi 221005}
\author{M.K.~Singh}
\altaffiliation{Participating as a member of TEXONO Collaboration}
\affiliation{Institute of Physics, Academia Sinica, Taipei 11529}
\affiliation{Department of Physics, Banaras Hindu University, Varanasi 221005}
\author{C.J.~Tang}
\affiliation{College of Physical Science and Technology, Sichuan University, Chengdu 610064}
\author{W.Y.~Tang}
\affiliation{Key Laboratory of Particle and Radiation Imaging (Ministry of Education) and Department of Engineering Physics, Tsinghua University, Beijing 100084}
\author{Y.~Tian}
\affiliation{Key Laboratory of Particle and Radiation Imaging (Ministry of Education) and Department of Engineering Physics, Tsinghua University, Beijing 100084}
\author{J.M.~Wang}
\affiliation{YaLong River Hydropower Development Company, Chengdu 610051}
\author{L.~Wang}
\affiliation{Key Laboratory of Particle and Radiation Imaging (Ministry of Education) and Department of Engineering Physics, Tsinghua University, Beijing 100084}
\author{Q.~Wang}
\affiliation{Key Laboratory of Particle and Radiation Imaging (Ministry of Education) and Department of Engineering Physics, Tsinghua University, Beijing 100084}
\author{Y.~Wang}
\affiliation{Key Laboratory of Particle and Radiation Imaging (Ministry of Education) and Department of Engineering Physics, Tsinghua University, Beijing 100084}
\author{S.Y.~Wu}
\affiliation{YaLong River Hydropower Development Company, Chengdu 610051}
\author{Y.C.~Wu}
\affiliation{Key Laboratory of Particle and Radiation Imaging (Ministry of Education) and Department of Engineering Physics, Tsinghua University, Beijing 100084}
\author{H.Y.~Xing}
\affiliation{College of Physical Science and Technology, Sichuan University, Chengdu 610064}
\author{Y.~Xu}
\affiliation{School of Physics, Nankai University, Tianjin 300071}
\author{T.~Xue}
\affiliation{Key Laboratory of Particle and Radiation Imaging (Ministry of Education) and Department of Engineering Physics, Tsinghua University, Beijing 100084}
\author{S.W.~Yang}
\altaffiliation{Participating as a member of TEXONO Collaboration}
\affiliation{Institute of Physics, Academia Sinica, Taipei 11529}
\author{N.~Yi}
\affiliation{Key Laboratory of Particle and Radiation Imaging (Ministry of Education) and Department of Engineering Physics, Tsinghua University, Beijing 100084}
\author{C.X.~Yu}
\affiliation{School of Physics, Nankai University, Tianjin 300071}
\author{H.J.~Yu}
\affiliation{NUCTECH Company, Beijing 100084}
\author{J.F.~Yue}
\affiliation{YaLong River Hydropower Development Company, Chengdu 610051}
\author{X.H.~Zeng}
\affiliation{YaLong River Hydropower Development Company, Chengdu 610051}
\author{M.~Zeng}
\affiliation{Key Laboratory of Particle and Radiation Imaging (Ministry of Education) and Department of Engineering Physics, Tsinghua University, Beijing 100084}
\author{Z.~Zeng}
\affiliation{Key Laboratory of Particle and Radiation Imaging (Ministry of Education) and Department of Engineering Physics, Tsinghua University, Beijing 100084}
\author{Y.H.~Zhang}
\affiliation{YaLong River Hydropower Development Company, Chengdu 610051}
\author{M.G.~Zhao}
\affiliation{School of Physics, Nankai University, Tianjin 300071}
\author{W.~Zhao}
\affiliation{Key Laboratory of Particle and Radiation Imaging (Ministry of Education) and Department of Engineering Physics, Tsinghua University, Beijing 100084}
\author{J.F.~Zhou}
\affiliation{YaLong River Hydropower Development Company, Chengdu 610051}
\author{Z.Y.~Zhou}
\affiliation{Department of Nuclear Physics, China Institute of Atomic Energy, Beijing 102413}
\author{J.J.~Zhu}
\affiliation{College of Physical Science and Technology, Sichuan University, Chengdu 610064}
\author{Z.H.~Zhu}
\affiliation{YaLong River Hydropower Development Company, Chengdu 610051}

\collaboration{CDEX Collaboration}
\noaffiliation



\date{\today}

\begin{abstract}
We report results of a search for light weakly interacting massive particle (WIMP) dark matter from the CDEX-1 experiment at the China Jinping Underground Laboratory (CJPL). Constraints on WIMP-nucleon spin-independent (SI) and spin-dependent (SD) couplings are derived with a physics threshold of 160 eVee, from an exposure of 737.1 kg-days. The SI and SD limits extend the lower reach of light WIMPs to 2 GeV and improve over our earlier bounds at WIMP mass less than 6 GeV.
\end{abstract}


\maketitle

\section{Introduction}

Compelling evidence from astroparticle physics and cosmology indicates that dark matter constitutes about 27\% of the energy density of our Universe~\cite{lab1}. WIMPs are the leading candidate for cold dark matter~\cite{lab2}, which is one kind of dark matter component.

With excellent energy resolution and low threshold, a kind of {\it p}-type point contact germanium detector ({\it p}PCGe)~\cite{lab3,lab4} has been used and further developed to search for light WIMPs in recent years. Earlier measurements of the phase-I experiment of the CDEX collaboration (CDEX-1)~\cite{lab4,lab5,lab6} have provided results on low-mass WIMP searches with a {\it p}PCGe detector of target mass 994 g, and the physics threshold reached 475 eVee (``eVee" represents electron equivalent energy). For CDEX, lower energy threshold and lower background level are two main directions to pursue to get more sensitive results on low mass dark matter searches.

Focused on the lower threshold, a new 1-kg-scale {\it p}PCGe has been designed and customized (named ``CDEX-1B" in this report) based on our first prototype detector used in CDEX-1. The point electrode contact was further redesigned to get lower detector capacitance. A junction field-effect transistor (JFET) with lower noise was selected to be used in an updated pre-amplifier specially designed for CDEX-1B. The supporting structure inside the cryostat was also modified to further suppress the background level.

Details of the CDEX-1B experimental setup, data analysis, and the constraints on WIMP-nucleon elastic scattering based on an exposure of 737.1 kg-days at CJPL are discussed in the subsequent sections.

\section{Experimental Setup}

\begin{figure}[!htbp]
  \includegraphics[width=1.0\linewidth]{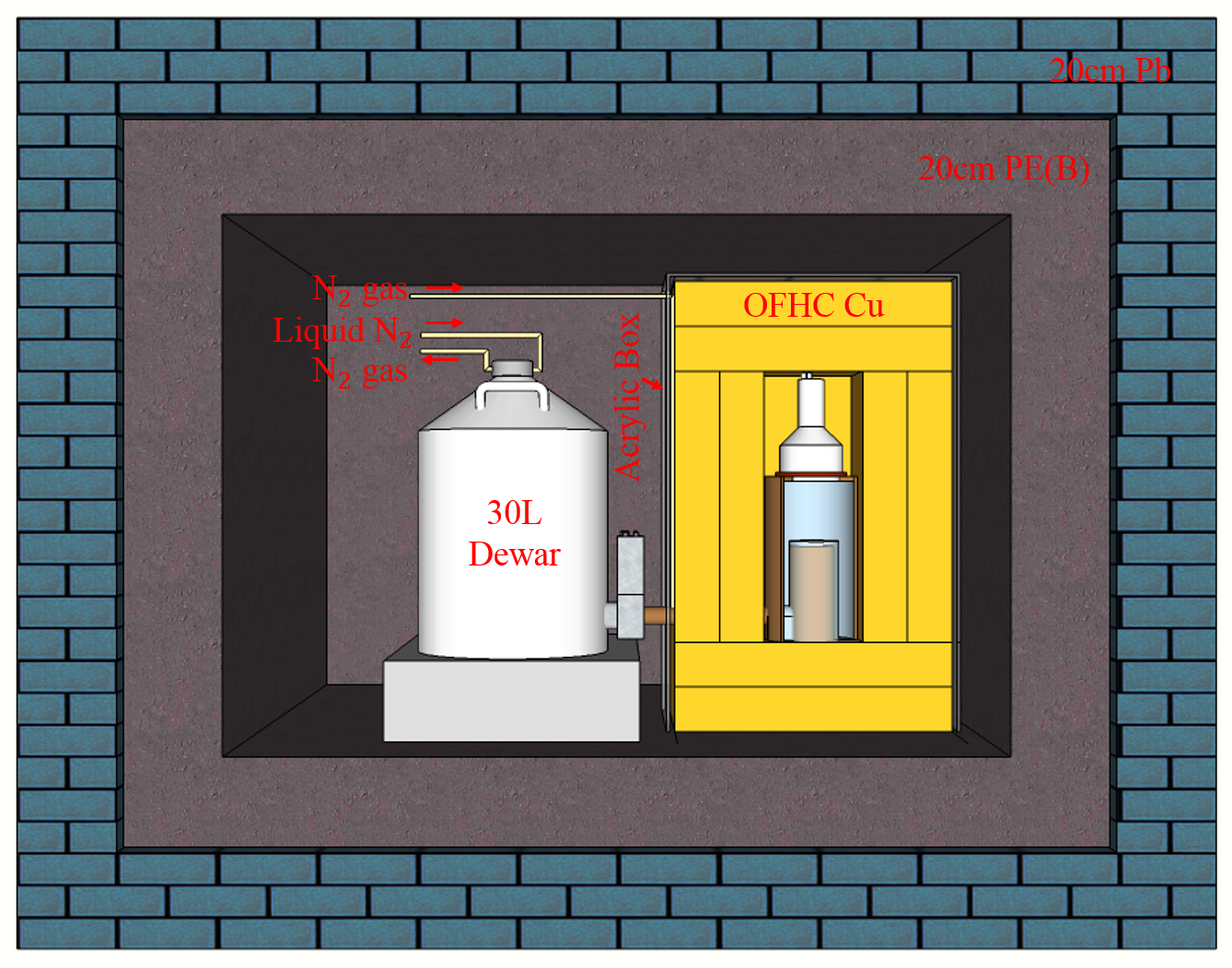}
  \caption{\label{fig:facility} Schematic diagram of CDEX-1B experimental setup, including the germanium (Ge) detector and NaI(Tl) anti-Compton detector, as well as the enclosing acrylic box and other passive shielding, including oxygen-free high conductivity copper, borated polyethylene and lead. The entire structure is placed inside a polyethylene room with walls of thickness 1 m, as described in Ref.~\cite{lab7}.}
\end{figure}

The shielding configuration of CDEX-1B is depicted in Fig.~\ref{fig:facility}. The germanium (Ge) cryostat and NaI(Tl) anti-Compton detectors are inside an oxygen-free high conductivity (OFHC) copper shielding with a thickness of 20 cm, then enclosed by an acrylic box, which was purged by nitrogen gas evaporated from the liquid nitrogen dewar. Other passive shielding includes, from inside to outside, 20 cm of borated polyethylene, 20 cm of lead and 1 m of polyethylene~\cite{lab4}. Detailed information about the passive shielding system can be found in Ref.~\cite{lab7}.

\begin{figure*}[!htbp]
  \includegraphics[width=1.0\linewidth]{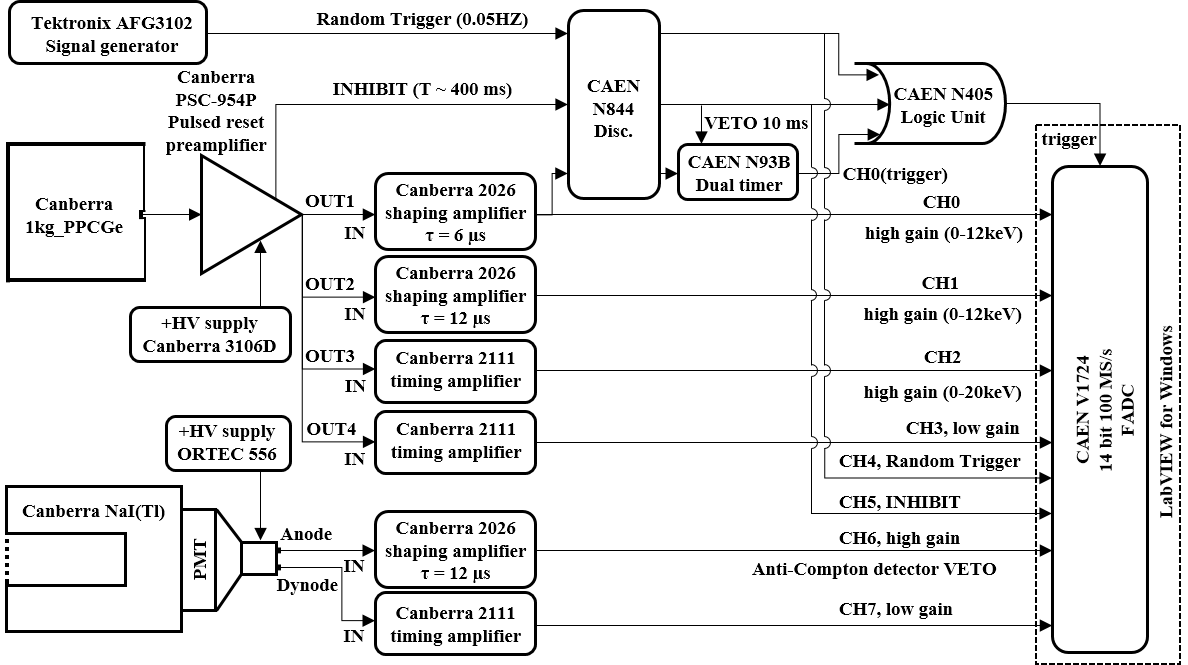}
  \caption{\label{fig:daq} Schematic diagram of the electronics and the DAQ system of the germanium detector and the NaI(Tl) detector.}
\end{figure*}

A schematic diagram of the electronics and data acquisition (DAQ) system is shown in Fig.~\ref{fig:daq}. The signals were read out by a pulsed reset preamplifier from the {\it p}$^{+}$ electrode of the {\it p}PCGe, and the preamplifier has four identical energy-related outputs. Two of them were distributed into shaping amplifiers with high gain for the low energy region (0-12 keV) at different shaping times, 6 $\mu$s (SA$_{6\mu s}$) and 12 $\mu$s (SA$_{12\mu s}$), respectively. The other two outputs were loaded to a timing amplifier (TA) which keeps relatively accurate time information. One with high gain for the medium energy region (0-20 keV) supplied the charge collection time. The other, with low gain for the high energy region (0-1.3 MeV), can be used to analyze the background origin.

Signals from the amplifiers were sampled and recorded by a 100 MHz flash analog-to-digital convertor (FADC). The recording time intervals were 120 $\mu$s for each channel. The output of SA$_{6\mu s}$ was fed into a leading-edge discriminator to supply the trigger of the DAQ system. In addition to that, the discriminator outputs of random trigger events at 0.05 Hz and the reset inhibit signal also served as triggers and were digitized. These random trigger (RT) events will be used to estimate the dead time of the DAQ system and derive the efficiencies of those analysis selection procedures which are uncorrelated with the pulse shape of the signal from the Ge detector. The reset inhibit signal recorded the discharging time, with a reset period of $\sim$400~ms, and a veto interval of 10 ms was applied after every reset to reject electronic-induced noise. The photomultiplier tube (PMT) outputs from the NaI(Tl) anti-Compton detector at two different gain factors were also digitized.

The germanium crystal mass of CDEX-1B is 1008 g, almost identical with the crystal used in CDEX-1. The main performance parameters of CDEX-1B are measured and listed in Table~\ref{tab:crystal_performance}, compared with those from CDEX-1. The pedestal RMS achieves 31 eVee, while that for CDEX-1 is 55 eVee. The full width at half maximum (FWHM) of a pulser input is 80 eVee, while that for CDEX-1 is 130 eVee. The FWHM at 10.37 keVee X-rays peak from $^{68,71}$Ge is (177$\pm$3) eVee. With all these improvements, a physics analysis threshold of 160 eVee at 17\% signal selection efficiency is achieved, which is significantly lower than earlier achievement by CDEX-1.

Starting on March 27th, 2014, the background measurement of CDEX-1B ran for 786.3 days until July 2017, apart from calibrations by gamma sources, neutron sources, and dead layer measurement performed in August 2014, January 2015 and March 2016, respectively. The dead time ratio of the DAQ system remained stable and was $<$0.1\% as measured by RT events.

The optimal area from the SA$_{6\mu s}$ channel~\cite{lab6} was chosen to define the energy (E) for its excellent energy linearity at the low energy range. Energy calibration was achieved using the internal cosmogenic X-ray peaks: $^{68}$Ge (10.37 and 1.30 keVee), $^{68}$Ga (9.66 keVee), and $^{65}$Zn (8.98 keVee). The zero-energy was defined by the RT events. The linearity is so good that the deviation is less than 0.4$\%$.

\section{Data Analysis}

Due to the extremely small cross section between WIMPs and nucleons, WIMP-nucleon ($\chi$-N) interactions are characterized by being single-site events uncorrelated with other detector components, while having the same pulse shape due to genuine physical processes~\cite{lab8}. A series of data analysis criteria were adopted to select WIMP-nucleon events, and their corresponding signal efficiencies were measured. The details are as follows.

(1) Anti-Compton Veto (AC):
The charge drifts in the Ge detector for a certain amount of time, while the photon in the NaI(Tl) anti-Compton detector is detected without time delay. The time difference between the NaI(Tl) anti-Compton detector and the Ge detector can be used to select  coincidence (anti-coincidence) events, which are denoted as AC$^{+(-)}$, respectively. Due to the extremely low trigger rates, the AC$^-$ selection discriminates the physics events at a signal efficiency of close to 100\%, which can be measured by RT events. Physics events selected by the AC$^+$ tag will be used to optimize the following selection criteria and calculate those selection efficiencies.

\begin{table*}[!htbp]
\caption{\label{tab:crystal_performance} The performance parameters of CDEX-1B, compared with the detector used in CDEX-1~\cite{lab4,lab5,lab6}.}
\begin{ruledtabular}
\begin{tabular}{ccccccc}
 &pedestal&FWHM$^{a}$ of&energy at 50\% &physics analysis&combined signal\\
 crystal&RMS&pulser&trigger efficiency&threshold&efficiency at threshold\\
  \hline
  CDEX-1 & 55 eVee& 130 eVee& $246\pm2$ eVee& 475 eVee&80\%\\
  CDEX-1B & 31 eVee& 80 eVee& $126\pm2$ eVee& 160 eVee&17\%\\
\end{tabular}
\end{ruledtabular}
\leftline{\quad\quad$^{a}$Full width at half maximum (FWHM).}
\end{table*}

\begin{figure*}[!htbp]
  \includegraphics[width=0.5\linewidth,height=0.4\linewidth]{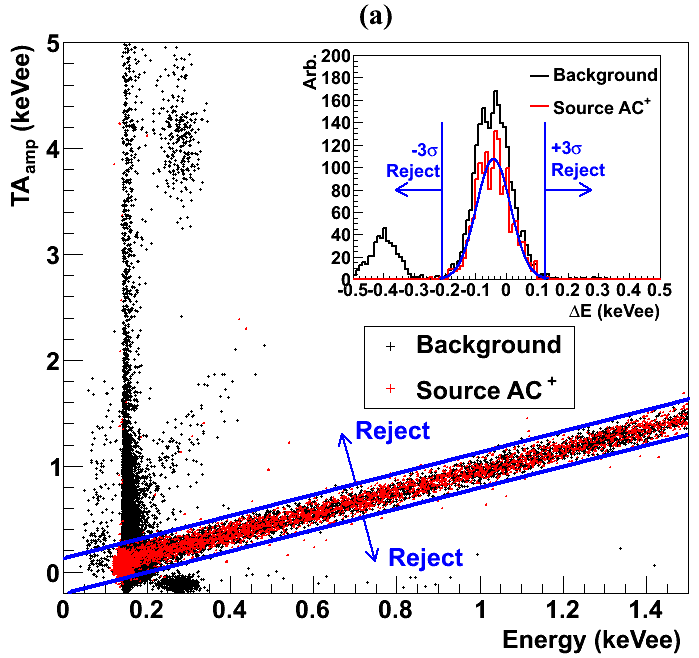}%
  \includegraphics[width=0.5\linewidth,height=0.4\linewidth]{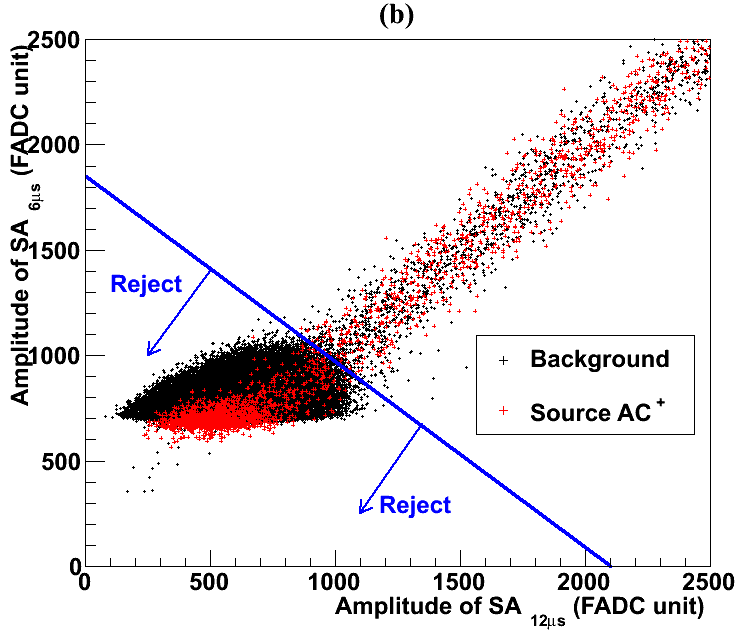}
  \caption{\label{fig:PSD} Two energy-dependent selections near energy threshold based on (a) correlations between the pulse height difference of TA channel (TA$_{amp}$) and calibrated energy; (b) correlations between the amplitude from SA$_{6\mu s}$ and SA$_{12\mu s}$ (PN).}
\end{figure*}

\newpage

\begin{figure}[!htbp]
  \includegraphics[width=1.0\linewidth]{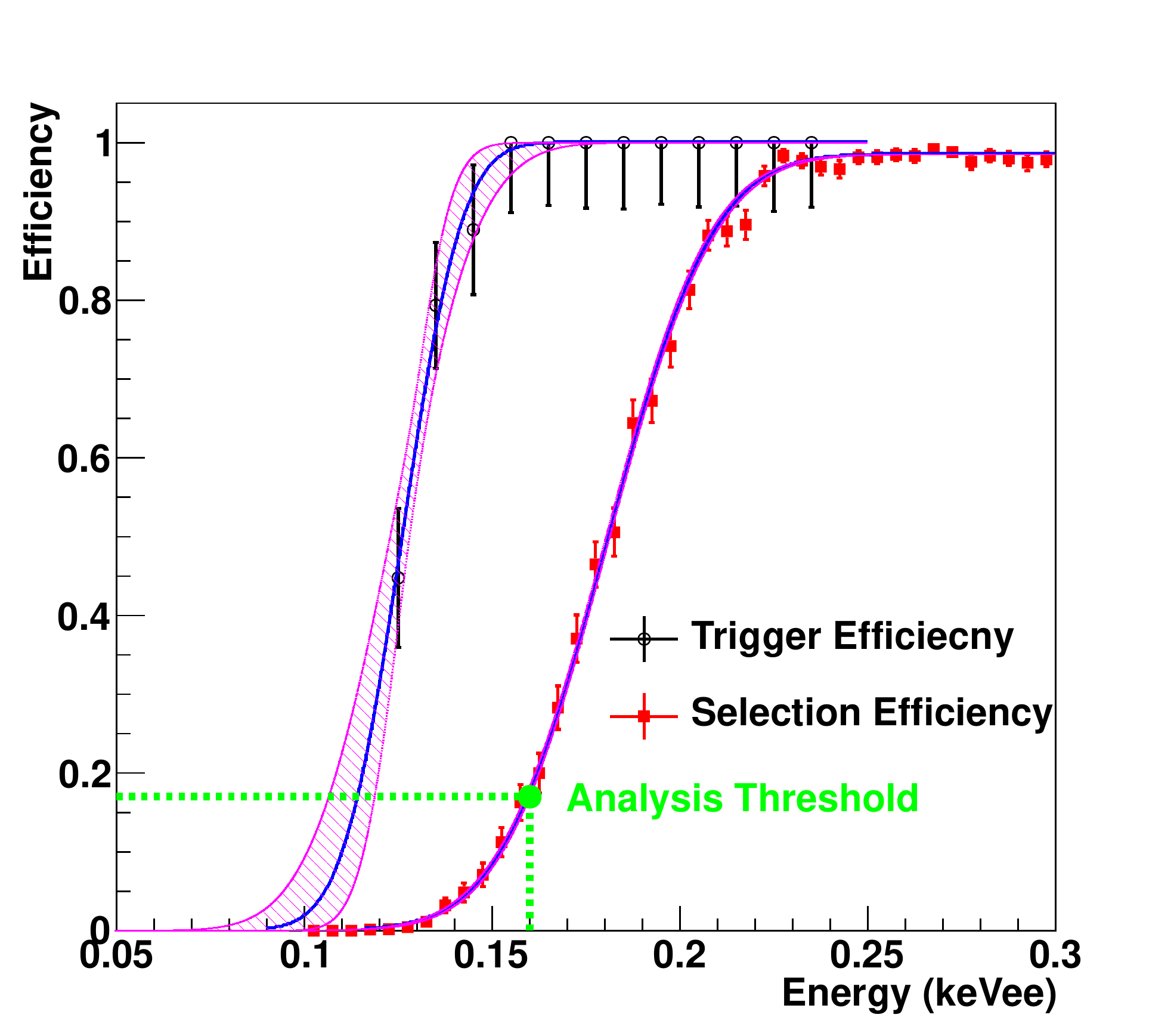}
  \caption{\label{fig:eff} Combined selection efficiencies and trigger efficiencies derived from source AC$^+$ events, fit with error functions.}
\end{figure}

(2) Basic Cuts (BC):
The energy correlations between TA$_{amp}$ and SA$_{6\mu s}$ are used to remove the surviving reset-induced noise, though a 10~ms gate after each reset inhibit signal has rejected the majority of that noise. TA$_{amp}$ is defined as the height difference between the maximum and the pedestal of the pulse from the TA channel. The selection criterion is that the energy difference ($\Delta$E) between TA$_{amp}$ and SA$_{6\mu s}$ cannot be larger than 3$\sigma$, which is derived from AC$^{+}$ events from gamma sources samples and ambient background, as displayed in Fig.~\ref{fig:PSD}(a). The signal selection efficiency is also derived from AC$^{+}$ events.

Cuts on the pedestals of SA$_{6\mu s,12\mu s}$ and TA (Ped) were also applied to discriminate those events whose pedestals exhibit anomalous behavior. Apart from using the minima (MIN) and the location of the maxima (t$\mathrm{_{MAX}}$) to remove abnormal events, the correlations between the amplitude from SA$_{6\mu s}$ and SA$_{12\mu s}$ (PN) were used to discriminate physics events from a large amount of noise near threshold, as displayed in Fig.~\ref{fig:PSD}(b). Detailed discussion of these selection cuts can be found in the literature~\cite{lab6,lab8}.

The selection efficiency of energy-independent cuts (Ped) is derived, from the survival of the RT events, to be 98.4\%. Selection efficiencies for energy-dependent selections (TA$_{amp}$, MIN, t$\mathrm{_{MAX}}$, PN) are from the survival of AC$^{+}$ events from source samples and {\it in situ} background. Figure~\ref{fig:eff} displays the combined BC selection efficiencies and trigger efficiencies calculated by source AC$^+$ events with red and black data points, respectively. The trigger efficiencies are defined as the fraction of the amplitude distribution above the threshold. For a certain amplitude of the signals, the optimal area from SA$_{6\mu s}$ has an approximately Gaussian distribution. The trigger efficiency will be approximately an error function, while the slope is related to the electronic noise level. The efficiency curves fitted with an error function, together with the 1$\sigma$ band of error, are also depicted in Fig.~\ref{fig:eff}, which shows a 50\% trigger threshold of 126$\pm$2~eVee. In this analysis, the physics analysis threshold is selected to be 160 eVee, which corresponds to a combined efficiency of 17$_{-0.7}^{+0.6}$\%.

\begin{figure}[!h]
  \includegraphics[width=\linewidth,height=0.7\linewidth]{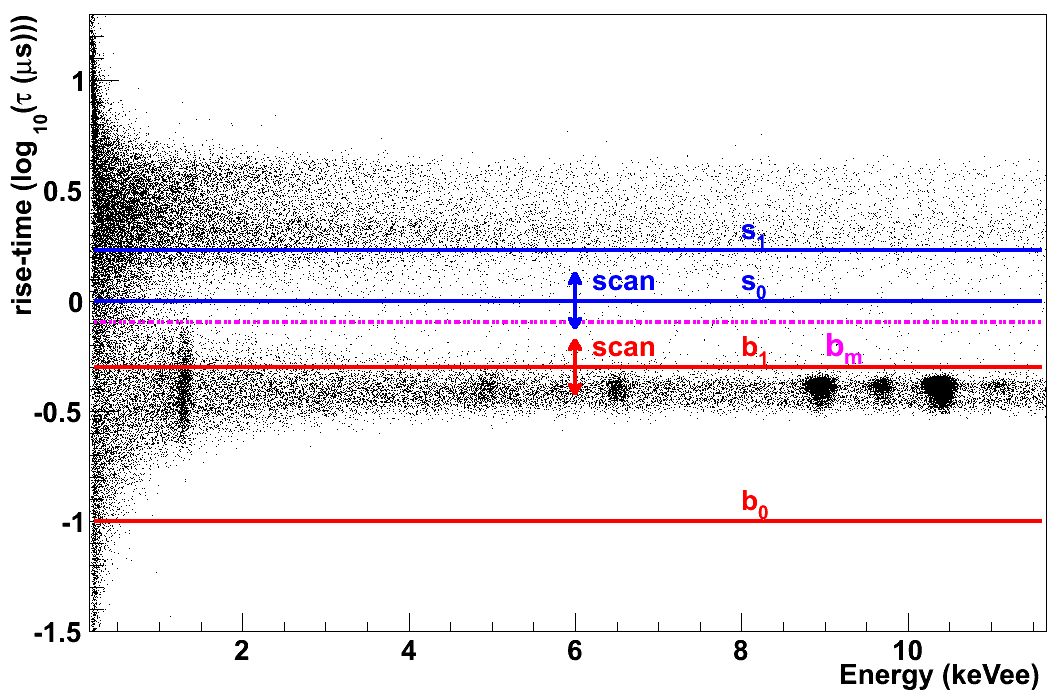}
  \caption{\label{fig:risetime} Rise-time versus energy scatter plot for the WIMP-induced candidate events based on AC$^{-}$ selection in CDEX-1B data.}
\end{figure}

(3) Bulk and Surface event selection (BS):
The outer {\it n}$^{+}$ dead layer of {\it p}PCGe is fabricated by lithium diffusion. The charge collection efficiency as a function of the depth of the surface was measured and simulated, showing that the thickness of the dead layer for the CDEX-1B detector is (0.88$\pm$0.12) mm~\cite{lab9}. This gives rise to a fiducial mass of 939 g and data exposure of 737.1 kg-days for this analysis.

Events depositing energy in the surface layer will have a partial charge collection with slower rise time ($\tau$) than those in the bulk volume. The rise time ($\tau$) is defined as the time interval between 5\% and 95\% of the TA pulse height, and is calculated by fitting to a hyperbolic tangent function to the TA pulse~\cite{lab5,lab6}. The scatter plot of log$_{10}[\tau]$ versus measured energy of AC$^{-}$ events from CDEX-1B is depicted in Fig.~\ref{fig:risetime}. The band structure characterizing bulk (B) and surface (S) events is well separated at energies larger than 2 keVee. The K-shell X-rays and L-shell X-rays from internal cosmogenic radioactivity are clearly identified. To get the corrected count rates of bulk events (B$_{r}$) at the low energy region of interest, the Ratio Method has been developed and detailed in a previous paper~\cite{lab10}.

The Ratio Method is based on the bulk/surface rise-time distribution PDFs (probability density functions) and the count ratios among different calibration sources:

\begin{figure*}[!htbp]
  \includegraphics[width=0.92\linewidth]{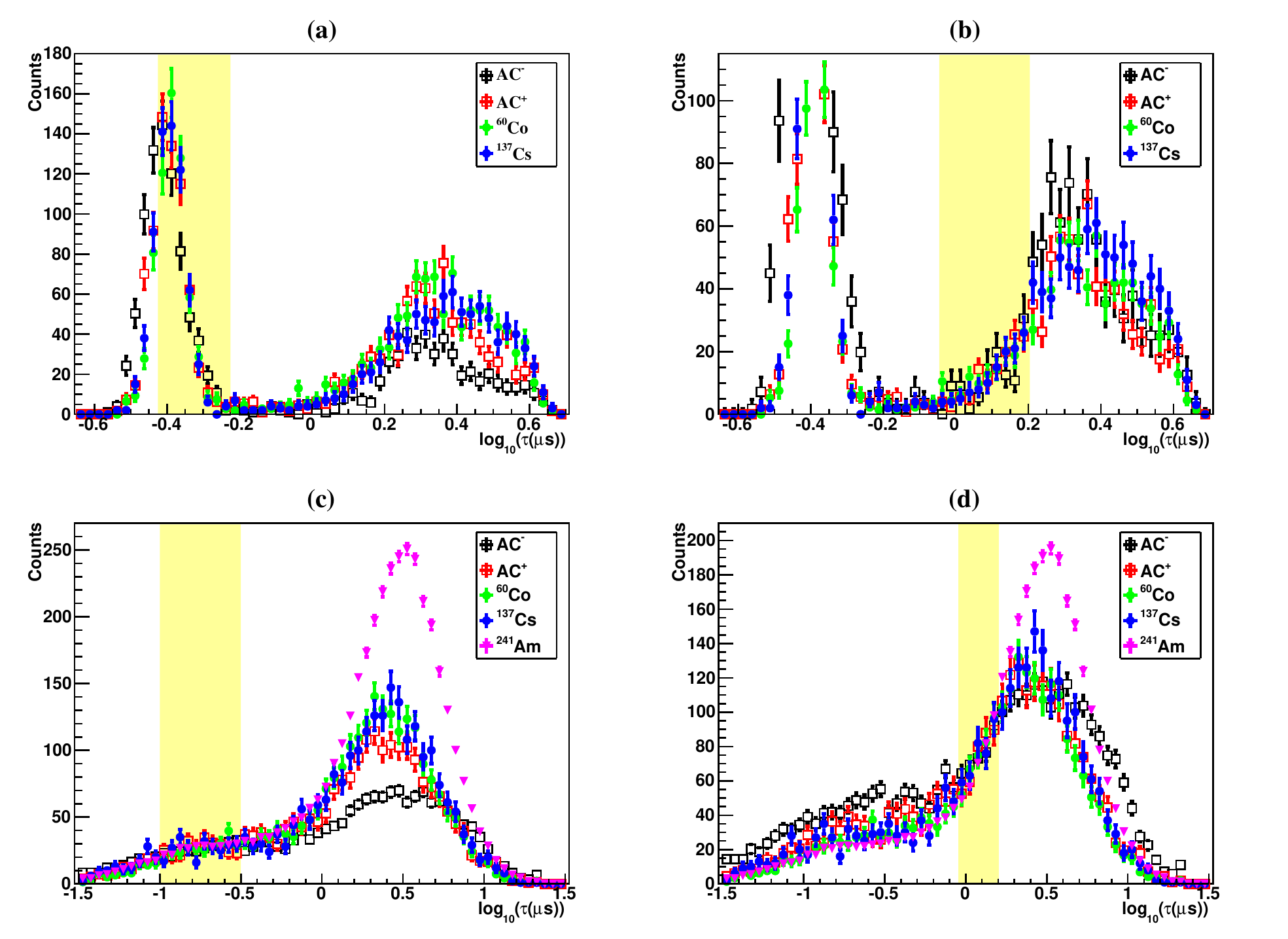}
  \caption{\label{fig:risetime_compare} Top: The rise-time distribution of input sources (AC$^-$, AC$^+$, $^{137}$Cs, $^{60}$Co) in the high energy region (5.0-5.5 keVee), with the shaded regions showing normalized intervals, (a) for bulk events (-0.425, -0.225) and (b) for surface events (log$_{10}$(0.9), log$_{10}$(1.6)). Bottom: The rise-time distribution of input sources (AC$^-$, AC$^+$, $^{137}$Cs, $^{60}$Co, $^{241}$Am) in the low energy region (0.2-0.4 keVee), with the shaded regions showing normalized intervals, (a) for bulk events (-1.0, -0.5) and (b) for surface events (log$_{10}$(0.9), log$_{10}$(1.6)).}
\end{figure*}

\begin{figure*}[!htbp]
  \includegraphics[width=0.92\linewidth]{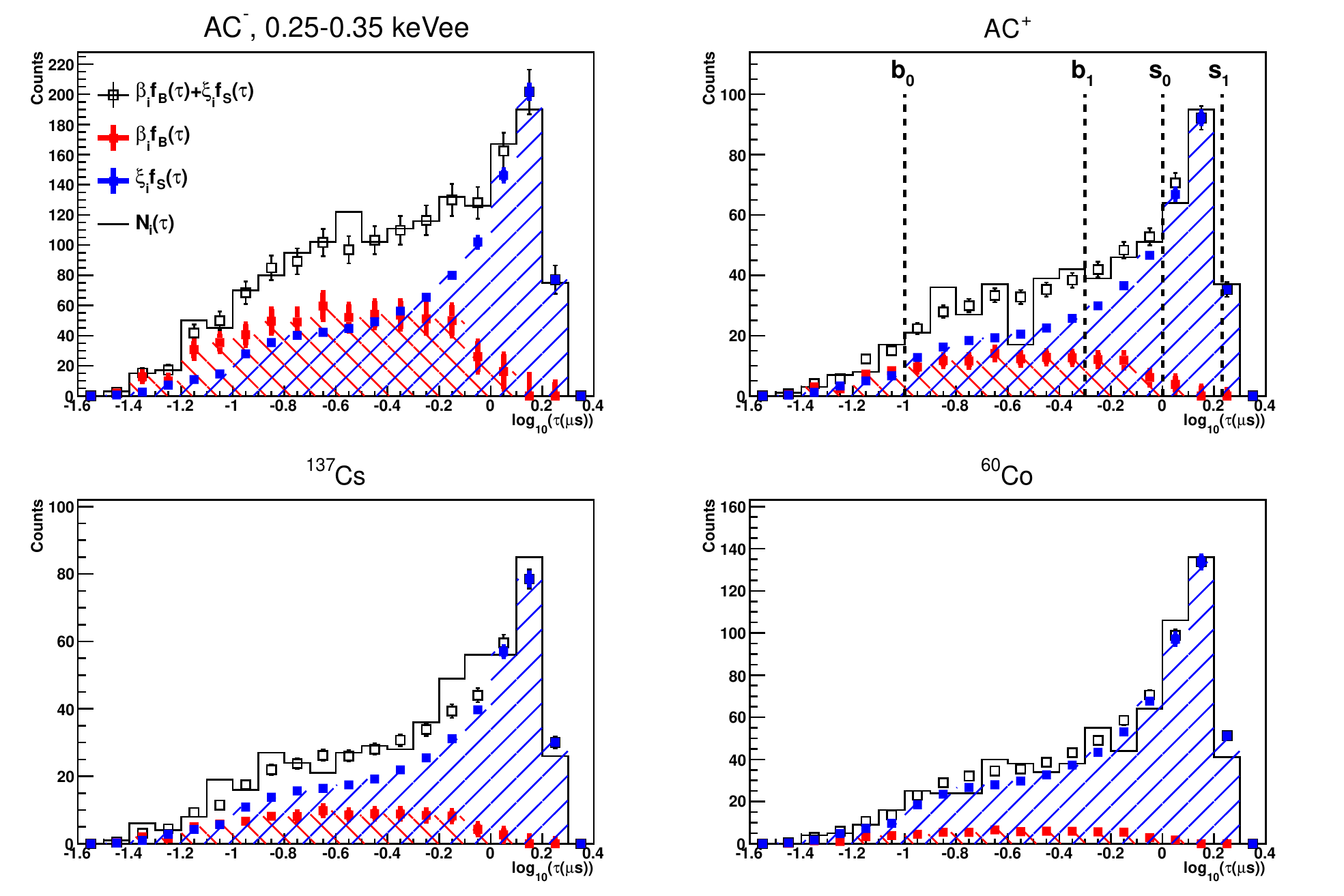}
  \caption{\label{fig:pdf_result} $f_B$ and $f_S$ distributions for the input sources (AC$^-$, AC$^+$, $^{137}$Cs, $^{60}$Co) at energy bins of 0.25-0.35 keVee. $^{241}$Am is also one input source, which supplies the $f_S$ distributions as a pure surface source.}
\end{figure*}

i) The validity of Ratio Method analysis requires that the calibration source data should have common bulk/surface rise-time distribution PDFs, and this condition has been checked and confirmed for input sources (AC$^-$, AC$^+$, $^{137}$Cs, $^{60}$Co) both at high energy region and low energy region. Figure~\ref{fig:risetime_compare} depicts the results of E=5.0-5.5 keVee and E=0.2-0.4 keVee. Bulk rise-time distributions are consistent among different sources. The measured surface rise-time distributions are different at high energy due to varying degrees of attenuation by the surface thickness. At near-threshold energies, however, the surface distributions are dominated by resolution smearing and are consistent among the various data samples. To avoid excessive estimation of B$_{r}$, low-energy $\gamma$'s from $^{241}$Am, which have been heavily attenuated by the surface thickness, can be adopted as calibration PDFs for S-samples only.

ii) Those input sources share the same PDFs but different scaling factors, which are proportional to the count ratios of each. The measured count rate of the $i^{th}$-source, $N_{i}$, as functions of $E$ and $\tau$ can be written as
\begin{small}
\begin{eqnarray}
N_{i}(E,\tau)&=&  N_{B i}(E,\tau)+N_{S i}(E,\tau) \nonumber\\
 &=& \beta_{i}(E)f_{B}(E,\tau)+\xi_{i}(E)f_{S}(E,\tau).
\label{eq:N_NB_NS}
\end{eqnarray}
\end{small}
where $N_{B i}$ and $N_{S i}$ denote the $(E,\tau)$-dependent functions of bulk and surface event distributions, respectively. $f_{B}(E,\tau)$ and $f_{S}(E,\tau)$ are the common PDFs shared by various input sources. $\beta_{i} (E)$ and $\xi_{i} (E)$ are the scaling factors, which are proportional to the count ratios. It is a special case that $\beta_{i} (E)=0$ for $i$=$^{241}$Am, because $^{241}$Am supplies the surface PDFs at low energies as a pure source.

As shown in Fig.~\ref{fig:risetime} and Fig.~\ref{fig:risetime_compare}, the ``pure" bulk region [$b_{0}$, $b_{1}$] and ``pure" surface region [$s_{0}$, $s_{1}$] are assumed to derive the count rates of each input source, by integrating counts at those regions~\cite{lab10}. In order to ensure  consistency between the background and those sources used to implement calibration, [$s_{0}$, $s_{1}$] was chosen to be close to the bulk band. Events with $\tau>s_{1}$ are removed as very-surface events and will be added back to the surface counts after calibration. Optimized $\beta_{i}(E)f_{B}(E,\tau)$ and $\xi_{i}(E)f_{S}(E,\tau)$ are obtained by $\chi^2$ minimization of the right hand-side of Eq.~\ref{eq:N_NB_NS},
\begin{small}
\begin{eqnarray}
\chi^{2}(E,\tau)&=&\sum_{i}\frac{[\beta_{i}(E)f_{B}(E,\tau)+\xi_{i}(E)f_{S}(E,\tau)-N_{i}(E,\tau)]^{2}}{\Delta{N}_{i}(E,\tau)^2}.\nonumber\\
&&
\label{eq:chi2_fB_fS}
\end{eqnarray}
\end{small}

iii) In the low energy range, the ``pure" region is only approximately pure, and the initial ratio value should be corrected by the PDFs obtained from the previous step. After $n$ iterations, for large $n$, convergent PDFs and ratios will be obtained. The best-fit results of $f_B(E,\tau)$ and $f_S(E,\tau)$ at E=0.25-0.35 keVee for CDEX-1B data are depicted in Fig.~\ref{fig:pdf_result}, and the real bulk/surface counts (B$_{r}$ and S$_{r}$) are calculated by integrating the bulk/surface PDFs. There is no definition of BS cut efficiency in the Ratio Method.
\begin{small}
\begin{eqnarray}
\beta^{(n)}_{i} (E) &=& \beta^{0}_{i}  (E) - \int_{b_{0}}^{b_{1}}{\xi^{(n-1)}_{i} (E)  ~ f^{(n-1)}_{S}(E, \tau) ~ d\tau},\nonumber\\
B_{ri}(E)&=&\int_{all~\tau}N_{Bi}(E,\tau)~d\tau\nonumber\\
&=&\int_{all~\tau}\beta^{(n)}_{i}(E)~f^{(n)}_{B}(E,\tau)~d\tau,
\label{eq:Br_Sr_calculation}
\end{eqnarray}
\end{small}
where $\beta^{0}_{i}(E)$ is the initial scaling factor given by integrating counts in the ``pure" bulk region, and $f^{(n-1)}_{B}(E,\tau)$ is the result of minimizing Eq.~\ref{eq:chi2_fB_fS} in the ${(n-1)}^{th}$-iteration, while $\xi^{(n)}_{i}(E)$ and $S_{r i}(E)$ for surface events have the same form as Eq.~\ref{eq:Br_Sr_calculation}.

The dashed line labelled $b_{m}$ (set as $\tau =$ 0.8 $\mu$s in this analysis) is used to roughly calculate the measured bulk count rate B$_{m}$, which should be corrected at low energy to get the real count rate B$_{r}$. Accordingly, the physics events are categorized by ``AC$^{-(+)}$$\otimes$B$_{r(m)}$" and ``AC$^{-(+)}$$\otimes$S$_{r(m)}$", where the superscript ${-(+)}$ denotes anti-coincidence(coincidence) with the Ge signals, and B$_r(m)$ or S$_r(m)$ denotes the bulk or surface count rate after(before) the BS correction. The values of B$_m$ and S$_m$ just give a reference to show the effectiveness of the Ratio Method in BS selection, and depend on the choice of selection criteria, which is set as $\tau<0.8~\mu s$ for B$_m$ (and $\tau>0.8~\mu s$ for S$_m$) in this analysis.

The main contributions to the total error of AC$^-$$\otimes$B$_{r}$ at threshold and at a typical high energy bin, are summarized in Table~\ref{tab:syserr}. Standard error propagation techniques are used to evaluate the combined uncertainties of B$_{r}$ and S$_{r}$. As discussed in Ref.~\cite{lab10}, the statistical errors mainly come from (a) errors at 1-$\sigma$ level on $f_B$ and $f_S$, calculated from $\chi_{min}^{2}+1$ for each $(E,\tau)$-bin; (b) errors of the initial scaling factors ($\beta_i^0$ and $\xi_i^0$), calculated from pure bulk/surface regions for every $E$-bin; (c) errors of correction terms of the final scaling factors after $(n-1)$ iterations ($\beta_i^{n-1}$ and $\xi_i^{n-1}$). The systematic uncertainties are evaluated by considering 4 terms: (a) the range of the ``pure" bulk/surface region ($[b_0,b_1]$ and $[s_0, s_1]$) is shifted; (b) some B/S calibration sources are removed; (c) rise-time distributions of some sources are shifted deliberately; and (d) the $\tau$ bin-size is changed.

\begin{table*}[!htb]
\begin{ruledtabular}
\caption{\label{tab:syserr} The main contributions to the total error of AC$^-$$\otimes$B$_{r}$ at threshold and at a typical high energy bin.}
\centering
\begin{tabular}{lcc}
Energy Bin & 0.16-0.21 keVee & 1.96-2.01 keVee \\
AC$^-$$\otimes$B$_r$ and Errors & $ 8.38 \pm 1.41 [$\rm stat$] \pm 2.19 [$\rm sys$]$ & $2.93 \pm 0.36 [$\rm stat$] \pm 0.17 [$\rm sys$]$ \\
(kg$^{-1}$keV$^{-1}$day$^{-1}$) & $=8.38 \pm 2.61 $ & $=2.93 \pm 0.40 $ \\
\hline
I) Statistical Uncertainties : & 1.41 & 0.36\\
\hline
II) Systematic Uncertainties : \\
    \hspace{0.8cm}(i) Choice of [b$_{0}$, b$_{1}$], [s$_{0}$, s$_{1}$] & \hspace{0cm}1.09 & \hspace{0cm}0.13 \\
    \hspace{0.8cm}(ii) Choice of sources & \hspace{0cm}1.18 & \hspace{0cm}0.05 \\
    \hspace{0.8cm}(iii) Shift of $\tau$ by 0.02 (log$_{10}$($\mu s$)) & \hspace{0cm}1.22 & \hspace{0cm}0.09 \\
    \hspace{0.8cm}(iv) $\tau$ bin-size & \hspace{0cm}0.87 & \hspace{0cm}0.03 \\
    \hspace{0.5cm}Combined : & 2.19 & 0.17 \\
\end{tabular}
\end{ruledtabular}
\end{table*}

\begin{figure*}[!htbp]
  \includegraphics[width=0.5\linewidth]{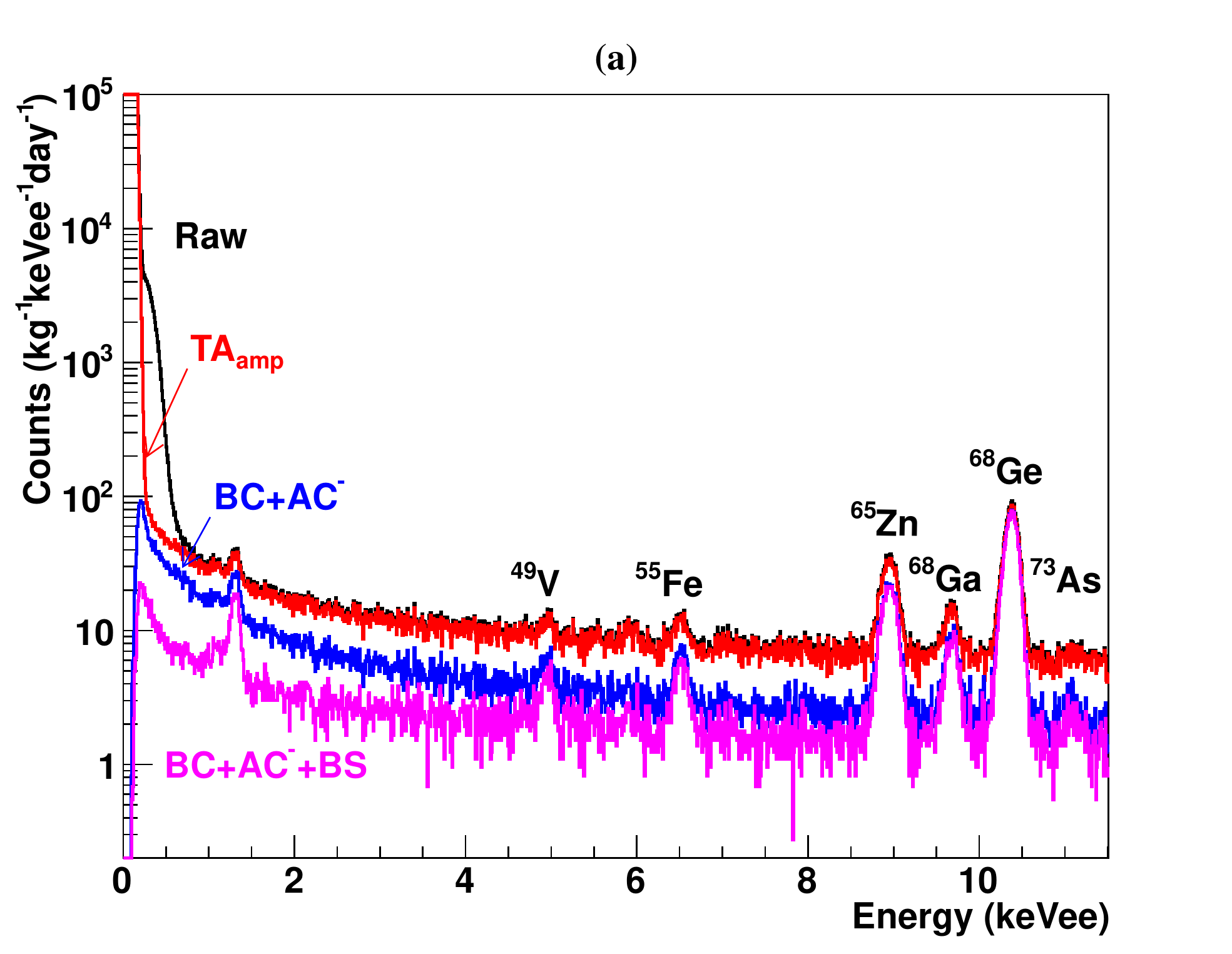}~\quad
  \includegraphics[width=0.5\linewidth]{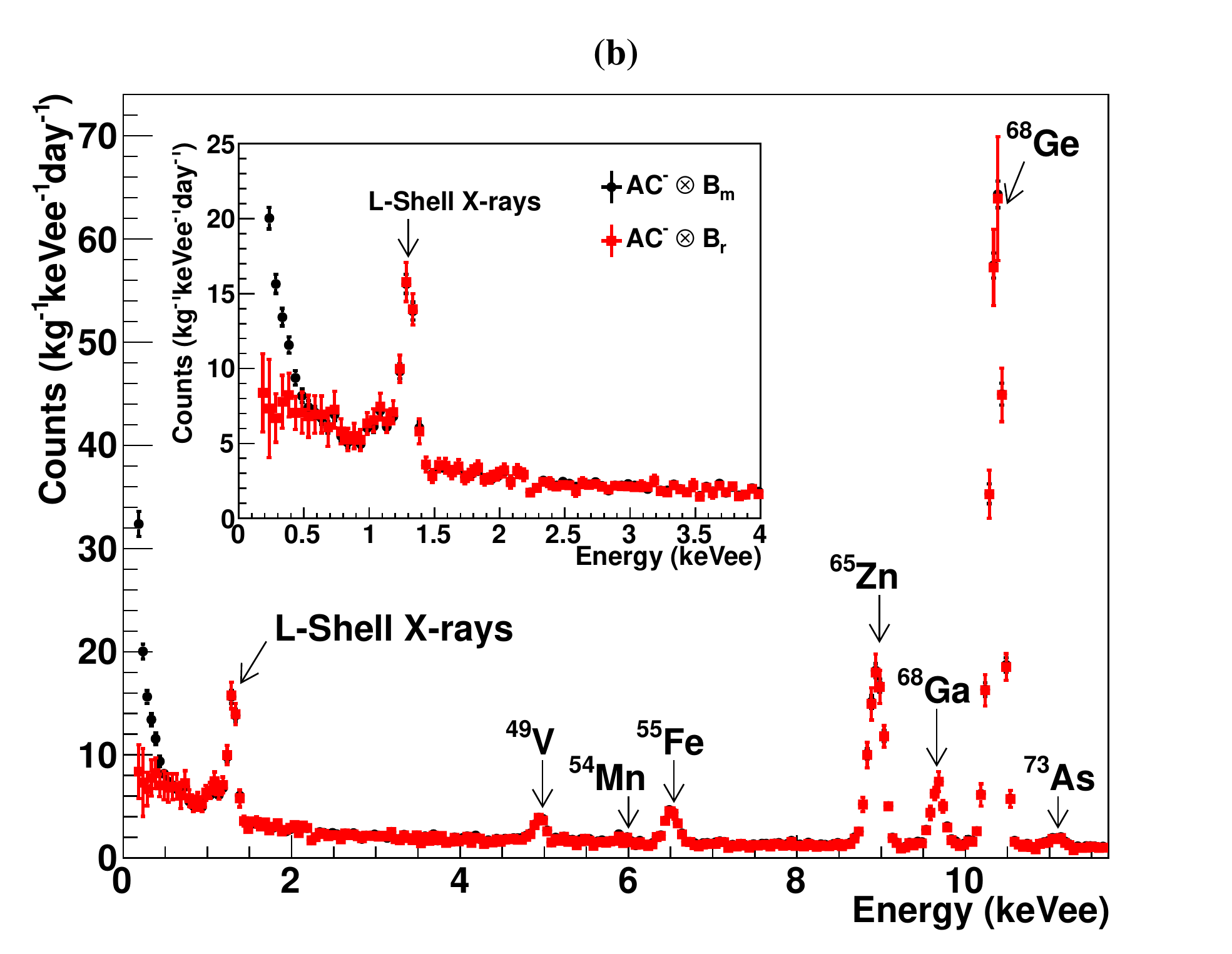}
  \includegraphics[width=0.5\linewidth]{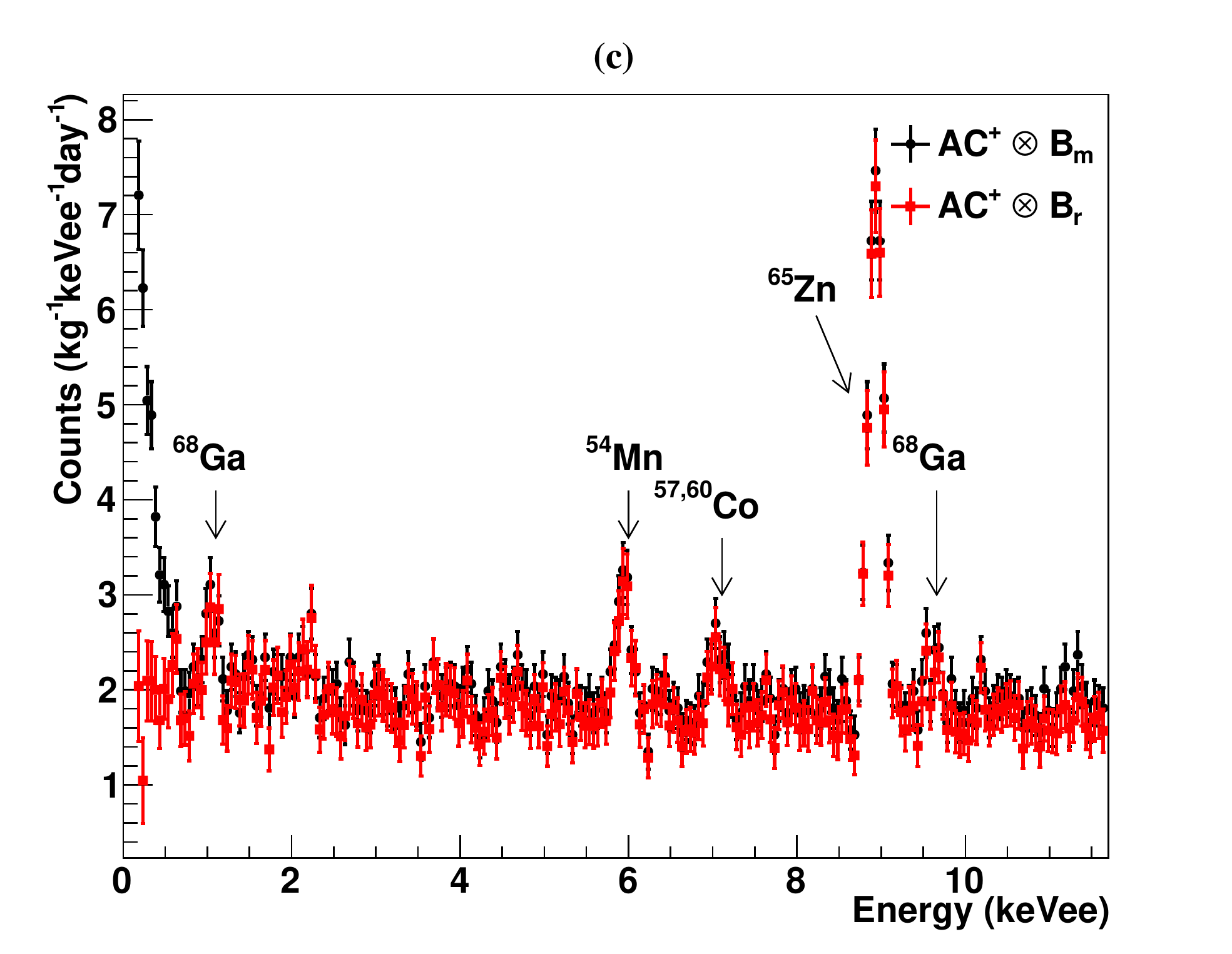}~\quad
  \includegraphics[width=0.5\linewidth,height=0.37\linewidth]{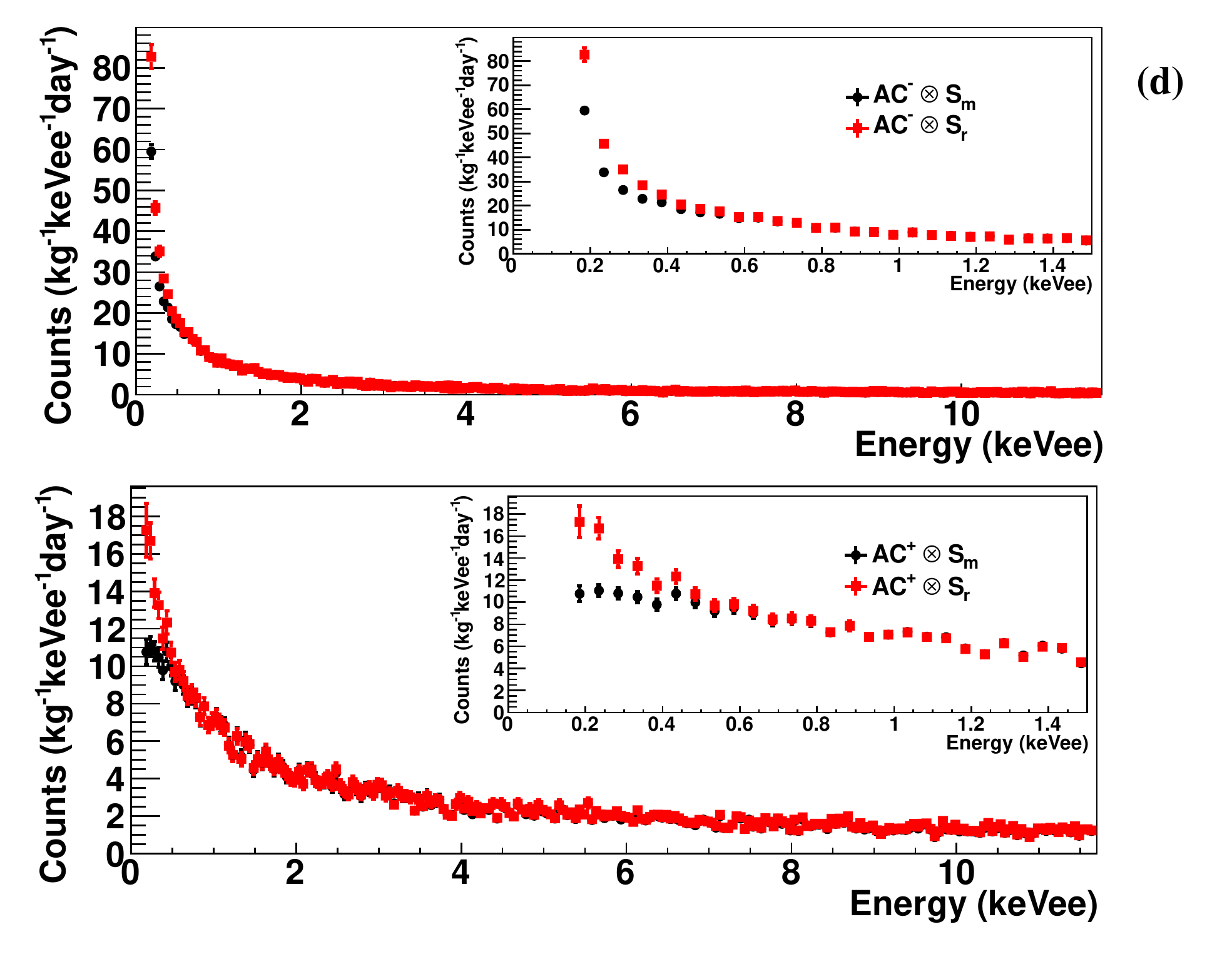}
  \caption{\label{fig:spec} (a) Measured energy spectra of CDEX-1B, showing the raw spectra and those at different stages of the analysis; (b) constructed AC$^{-}\otimes$B$_{r}$ spectra and the measured AC$^{-}\otimes$B$_{m}$ ($\tau <$ 0.8 $\mu$s) spectra; (c) constructed AC$^{+}\otimes$B$_{r}$ spectra and the measured AC$^{+}\otimes$B$_{m}$ spectra, with some X-ray peaks also identified; and (d) top: AC$^{-}\otimes$S$_{r}$ and AC$^{-}\otimes$S$_{m}$ spectra; bottom: AC$^{+}\otimes$S$_{r}$ and AC$^{+}\otimes$S$_{m}$ spectra.}
\end{figure*}

The measured raw spectra and those at different stages of the analysis are depicted in Fig.~\ref{fig:spec}(a) without considering the efficiency corrections. Combining the systematic and statistical errors, the constructed AC$^{-}\otimes$B$_{r}$ spectrum is depicted in Fig.~\ref{fig:spec}(b), as well as the measured AC$^{-}\otimes$B$_{m}$ ($\tau <$ 0.8 $\mu$s) spectrum. The trigger efficiency and cut efficiency have been considered. The minimum energy is at 160 eVee, matching the first finite efficiency bin of Fig.~\ref{fig:eff}. Several characteristic X-ray peaks from internal cosmogenic radioactive isotopes can be identified clearly, and include $^{68,71}$Ge, $^{68}$Ga, $^{73,74}$As, $^{65}$Zn, $^{55}$Fe, $^{54}$Mn, and $^{49}$V. As the characteristic X-rays are internal and short-ranged, the detection efficiency is almost 100\%. Figure~\ref{fig:spec}(c) displays the AC$^{+}\otimes$B$_{r}$ and AC$^{+}\otimes$B$_{m}$ spectra, in which X-rays peaks from $^{68}$Ga (9.66 keVee), $^{65}$Zn (8.98 keVee), $^{57,60}$Co (7.11 keVee), and $^{54}$Mn (5.99 keVee) can be identified clearly. These peaks are produced from  cascade decays corresponding to related cosmogenic radioactive isotopes, and hence a portion of these high energy gammas escape from the Ge detector and are tagged by the NaI(Tl) anti-Compton detector. This is a non-trivial demonstration of validity of this analysis, since every energy bin is processed independently of the others. The constructed AC$^{-}\otimes$S$_{r}$ and AC$^{+}\otimes$S$_{r}$ spectra are depicted in Fig.~\ref{fig:spec}(d), together with the measured AC$^{-}\otimes$S$_{m}$ and AC$^{+}\otimes$S$_{m}$ spectra.

\section{Residual Spectrum Analysis}

The ratios of the K-shell to L-shell X-ray events based on Ref.~\cite{lab15} are used to predict the intensity of the L-shell X-rays in the lower energy ranges ($<$1.6 keVee). In addition to the L-shell X-ray contributions, there are residual excess events in the AC$^{-}$ spectrum at low energy $<$3 keVee. Several candidate sources were examined but all fail to explain these anomalous events:

1) Cosmic rays: With a rock overburden of more than 2400 m giving rise to a measured muon flux of 61.7 y$^{-1}$m$^{-2}$~\cite{lab16} at CJPL, the contribution from muons can be neglected.

2) Neutrons: Simulation studies and measurements with a $^{232}$Cf neutron source show neutron-induced events in both AC$^{-}$ and AC$^{+}$ spectra. This excess is not observed in the AC$^{+}$ spectra. The thermal neutron flux around the copper shielding has been measured with a $^{3}$He tube, and the thermal neutron flux is (2.8$\pm$1.2)$\times$10$^{-8}$ cm$^{-2}$s$^{-1}$~\cite{lab17}. By simulations with Geant4~\cite{lab177}, the contributions from thermal neutrons can also be  neglected.

3) Tritium: Tritium ($^{3}$H) will produce a continuous beta energy spectrum with an end point energy of 18.6 keV. The spectrum should be almost flat below 2 keVee and will not contribute to the abnormal rising. From the cosmic ray simulation combined with background data analysis, the $^{3}$H count rate is 1.81 kg$^{-1}$day$^{-1}$ at energy range 0-18.6 keVee, far less than the current background count rate of 13.34 kg$^{-1}$day$^{-1}$.

Extensive tests related to the passivated surface, such as excess $^{210}$Pb or $^{210}$Po on the tin at the central contact or the signal readout brass bin, degraded beta emissions from $^{40}$K in PTFE, etc., have been carried out to explain those anomalous counts. These depend on the details of material components in the vicinity of the crystal, and research is ongoing.

\begin{figure*}[!htbp]
  \includegraphics[width=0.5\linewidth]{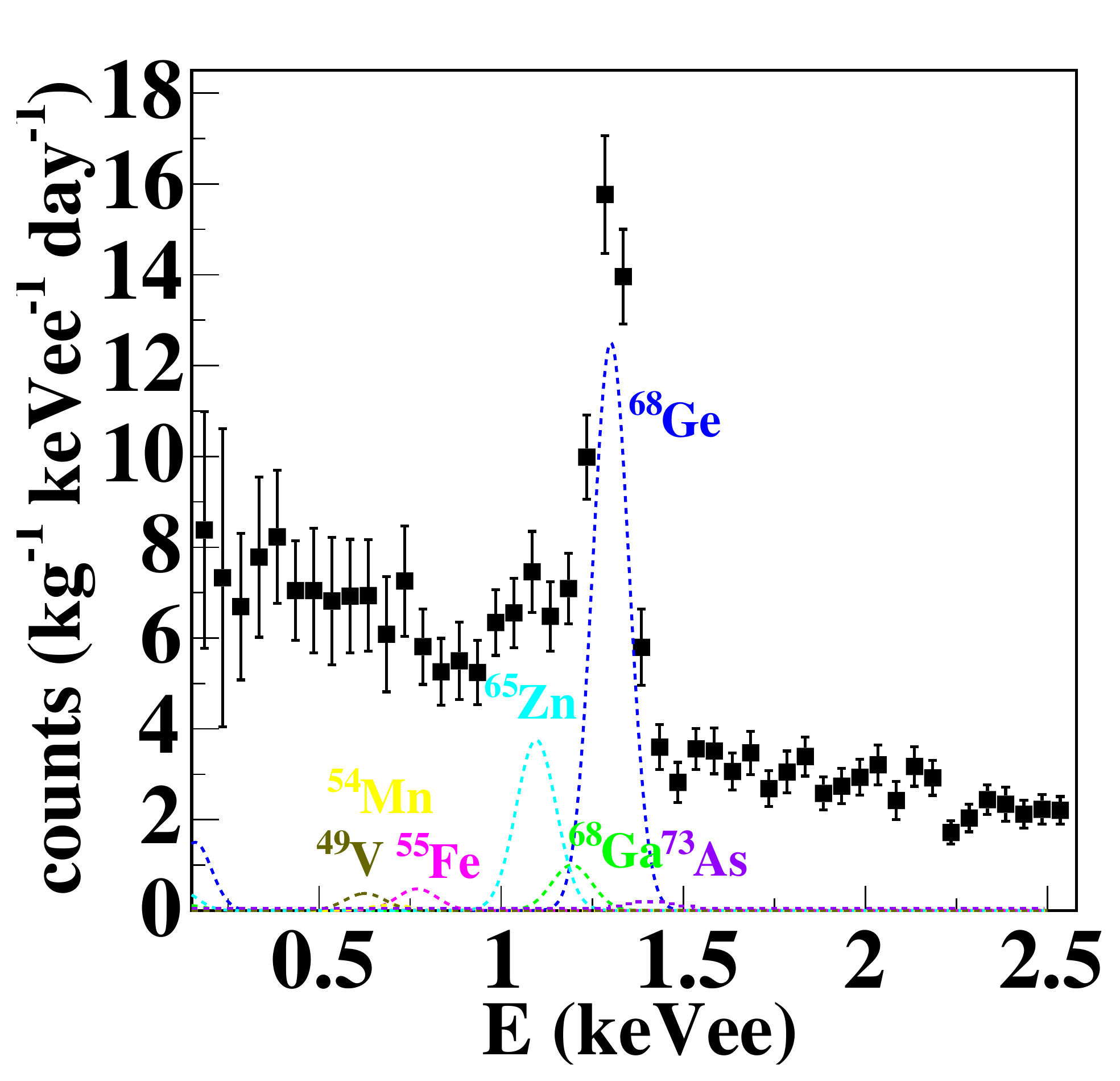}%
  \includegraphics[width=0.5\linewidth]{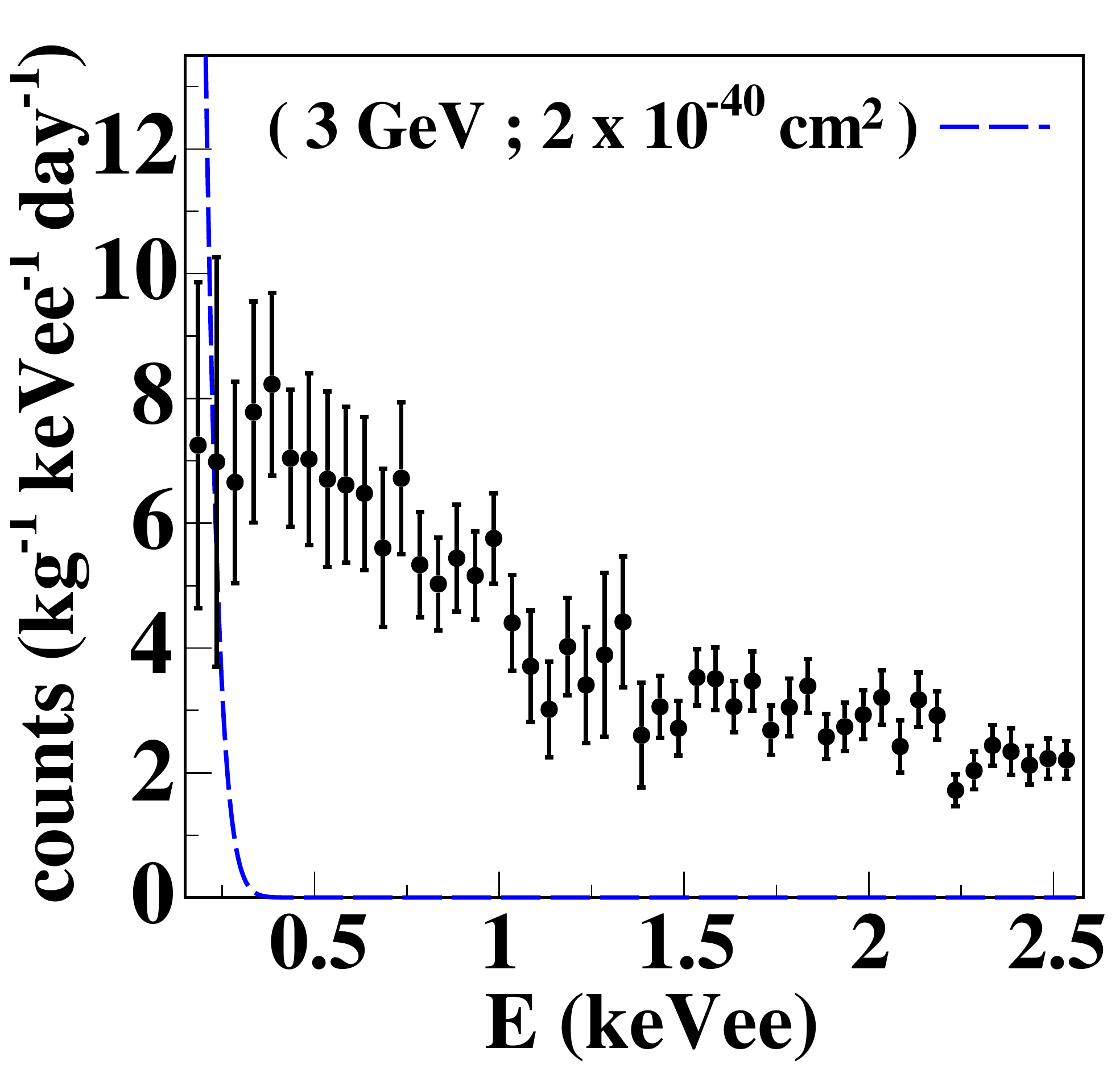}
  \caption{\label{fig:KL_Xrays} Left (a): The dashed lines displaying the contributions of the L/M-shell X-rays are derived from K-shell X-ray intensities. Right (b): The residual spectrum with L/M-shell X-ray contributions subtracted, together with the exclusion line at m$_{\chi} =$ 3 GeV/c$^{2}$ and spin-independent ${\chi}$-N cross section $\sigma^{SI}_{\chi N}=$ 2$\times$$10^{-40}$ cm$^{2}$ by the binned Poisson method~\cite{lab18}.}
\end{figure*}

\section{Dark Matter Constraints}

\begin{figure*}[!htbp]
  \includegraphics[width=0.5\linewidth,height=0.5\linewidth]{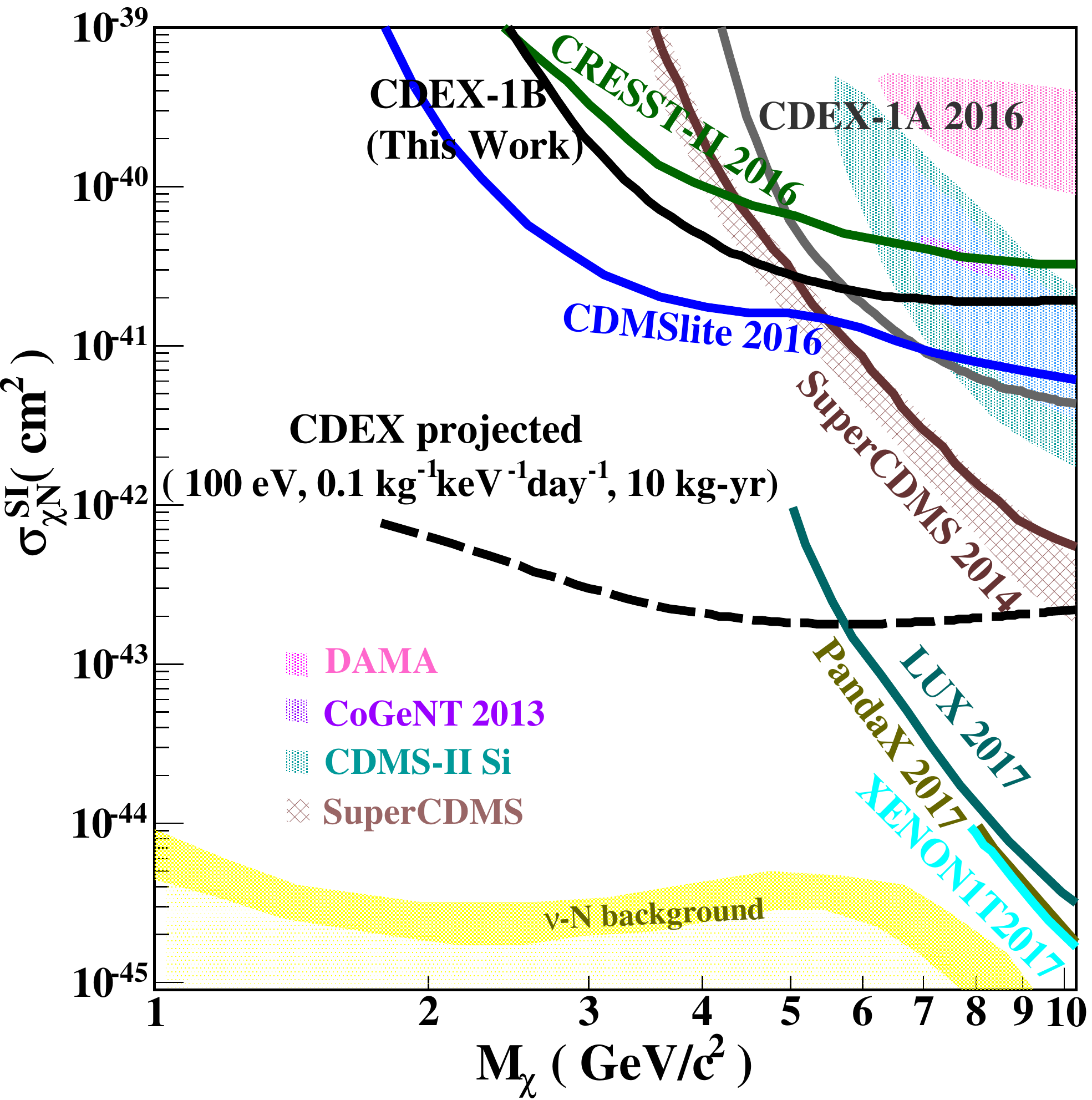}%
  \includegraphics[width=0.5\linewidth,height=0.5\linewidth]{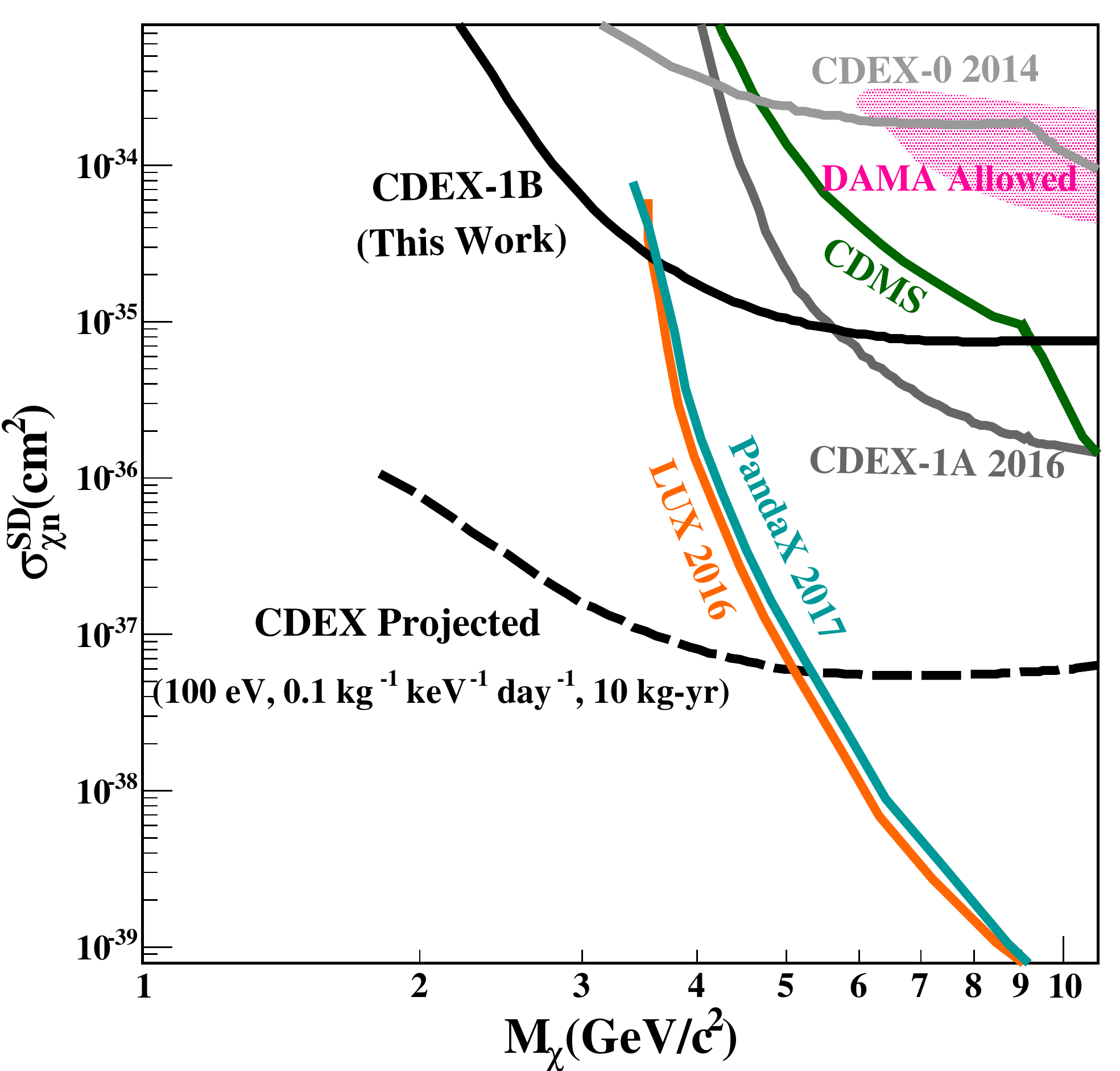}
  \caption{\label{fig:ex-plot} Left (a): Exclusion plot of spin-independent ${\chi}$-N coupling at 90\% confidence level, superimposed with results from other benchmark experiments. The CDEX-1 results from this work are depicted in solid black. Allowed regions given by CoGeNT~\cite{lab22}, DAMA/LIBRA~\cite{lab23} and CDMS-II (Si)~\cite{lab24} are presented, as well as the exclusion limits from CDEX-1~\cite{lab6}, XENON1T~\cite{lab25}, LUX~\cite{lab26}, PandaX-II~\cite{lab266}, CDMSlite~\cite{lab27}, SuperCDMS~\cite{lab28} and CRESST-II~\cite{lab29}. The ``Neutrino Floor" due to solar neutrino $\nu$-N scattering is also displayed~\cite{lab30}. Right (b): Exclusion plot of spin-dependent ${\chi}$-neutron coupling at 90\% confidence level, along with the allowed regions given by DAMA/LIBRA~\cite{lab23} and exclusion limits from  CDEX-1~\cite{lab6,lab31}, LUX~\cite{lab26}, PandaX-II~\cite{lab266}, and CDMS~\cite{lab28}. The potential reach with target sensitivities of 100 eVee threshold at 0.1~kg$^{-1}$ keV$^{-1}$ day$^{-1}$ background level for 10 kg-year exposure are also superimposed both for spin-independent and spin-dependent ${\chi}$N couplings.}
\end{figure*}

A conservative analysis with only the subtraction of the L/M-shell X-rays was performed. As shown in Fig.~\ref{fig:KL_Xrays}(a), the contributions of the L/M-shell X-rays are derived from K-shell X-ray intensities. The final residual spectrum with L/M-shell X-ray contributions subtracted in the region 0.16-2.5 keVee is shown in Fig.~\ref{fig:KL_Xrays}(b), together with the exclusion line at m$_{\chi} =$ 3 GeV/c$^{2}$ and the spin-independent ${\chi}$-N cross section $\sigma^{SI}_{\chi N} =$ 2$\times$$10^{-40}$ cm$^{2}$ by the binned Poisson method~\cite{lab18}, which is applied to this residual spectrum in this analysis. The quenching factor is provided by the TRIM program~\cite{lab19}, coupled with a 10\% systematic error implied by the spread of the measured data at the recoil energy of 254 eV to 10 keV. A standard WIMP galactic halo assumption~\cite{lab20} and conventional astrophysical models~\cite{lab21} are used, with a local WIMP density of 0.3~\text{GeV}/\text{cm}$^{3}$ and a Maxwellian velocity distribution with $\nu_{0} =$ 220 \text{km}/\text{s} and escape velocity $\nu_{\text{esc}} =$ 544$~$\text{km}/\text{s}. The energy resolution of the detector is derived from the {\it in situ} cosmogenic peaks.

The derived exclusion plots for the spin-independent $\chi$-N coupling (SI) at 90\% confidence level is shown in Fig.~\ref{fig:ex-plot}(a), in which several benchmark experiments~\cite{lab6,lab22,lab23,lab24,lab25,lab26,lab266,lab27,lab28,lab29} are superimposed. Constraints for spin-dependent $\chi$-neutron (SD) scattering at 90\% confidence level is shown in Fig.~\ref{fig:ex-plot}(b), along with the allowed regions given by DAMA/LIBRA~\cite{lab23} and exclusion limits from  CDEX-1~\cite{lab6,lab31}, LUX~\cite{lab26}, PandaX-II~\cite{lab266}, and CDMS~\cite{lab28}. These results extend the low reach of light WIMPs to 2 GeV, and improve over our earlier bounds on SI for ${\text{m}}_{\chi} <$ 6 GeV~\cite{lab6} and those from the CRESST-II experiment 2016~\cite{lab29}, while achieving the best sensitivity on SD for ${\text{m}}_{\chi} <$ 4 GeV.

\section{Summary and Prospects}

We report results from a 1~kg-scale {\it p}PCGe detector with a physics threshold of 160 eVee at CJPL. By improving electronics and design, the energy threshold and noise are greatly reduced, moving forward to the goal of detecting low mass dark matter. Conservative constraints on WIMP-nucleon spin-independent and spin-dependent elastic scattering are derived based on an exposure of 737.1 kg-days. These results improve over our earlier bounds at ${\text{m}}_{\chi} <$ 6 GeV~\cite{lab6} while extending the low reach of light WIMPs to 2 GeV, and achieve the best sensitivity on SD for ${\text{m}}_{\chi} <$ 4 GeV, which show the effectiveness of reducing the energy threshold.

The high-statistics data reveals anomalous features with the AC$^{-}$ spectra below 3 keVee. Analysis suggests its possible origin to be at the vicinity of the {\it p}$^{+}$ contact of the Ge-crystal. Such background can be suppressed through radioactivity screening of materials used in future detectors. The electronics will be further optimized to target lower thresholds. The potential reach with target sensitivities of 100 eVee threshold at 0.1~kg$^{-1}$ keV$^{-1}$ day$^{-1}$ background level for 10 kg-year exposure is superimposed in Fig.~\ref{fig:ex-plot}.




\begin{acknowledgments}
This work is supported by the National Key Research and Development Program of China (No. 2017YFA0402200 and 2017YFA0402201) and the National Natural Science Foundation of China (Nos.11175099, 11275107, 11475117, 11475099, 11475092 and 11675088) and the National Basic Research Program of China (973 Program) (2010CB833006) . We thank the support of grants from the Tsinghua University Initiative Scientific Research Program (No.20121088494, 20151080354) and the Academia Sinica Investigator Award 2011-15, contracts 103-2112-M-001-024 and 104-2112-M-001-038-MY3 from the Ministry of Science and Technology of Taiwan.
\end{acknowledgments}




\begin{thebibliography}{0}%
\makeatletter
\providecommand \@ifxundefined [1]{%
 \@ifx{#1\undefined}
}%
\providecommand \@ifnum [1]{%
 \ifnum #1\expandafter \@firstoftwo
 \else \expandafter \@secondoftwo
 \fi
}%
\providecommand \@ifx [1]{%
 \ifx #1\expandafter \@firstoftwo
 \else \expandafter \@secondoftwo
 \fi
}%
\providecommand \natexlab [1]{#1}%
\providecommand \enquote  [1]{``#1''}%
\providecommand \bibnamefont  [1]{#1}%
\providecommand \bibfnamefont [1]{#1}%
\providecommand \citenamefont [1]{#1}%
\providecommand \href@noop [0]{\@secondoftwo}%
\providecommand \href [0]{\begingroup \@sanitize@url \@href}%
\providecommand \@href[1]{\@@startlink{#1}\@@href}%
\providecommand \@@href[1]{\endgroup#1\@@endlink}%
\providecommand \@sanitize@url [0]{\catcode `\\12\catcode `\$12\catcode
  `\&12\catcode `\#12\catcode `\^12\catcode `\_12\catcode `\%12\relax}%
\providecommand \@@startlink[1]{}%
\providecommand \@@endlink[0]{}%
\providecommand \url  [0]{\begingroup\@sanitize@url \@url }%
\providecommand \@url [1]{\endgroup\@href {#1}{\urlprefix }}%
\providecommand \urlprefix  [0]{URL }%
\providecommand \Eprint [0]{\href }%
\providecommand \doibase [0]{http://dx.doi.org/}%
\providecommand \selectlanguage [0]{\@gobble}%
\providecommand \bibinfo  [0]{\@secondoftwo}%
\providecommand \bibfield  [0]{\@secondoftwo}%
\providecommand \translation [1]{[#1]}%
\providecommand \BibitemOpen [0]{}%
\providecommand \bibitemStop [0]{}%
\providecommand \bibitemNoStop [0]{.\EOS\space}%
\providecommand \EOS [0]{\spacefactor3000\relax}%
\providecommand \BibitemShut  [1]{\csname bibitem#1\endcsname}%
\let\auto@bib@innerbib\@empty
\end{thebibliography}%


\begin{thebibliography}{99}
\bibitem{lab1} J. Beringer et al., Phys. Rev. D, {\bf 86}: 010001 (2012) and references therein; P. A. R. Ade et al., A\&A, {\bf571}:A16 (2014)
\bibitem{lab2} C. Kelso, D. Hooper, and M. R. Buckley, Phys. Rev. D, {\bf 85}: 043515 (2012) and references therein
\bibitem{lab3} P. S. Barbeau et al., J. Cosmol. Astropart. Phys., {\bf 09}:009 (2007)
\bibitem{lab4} \bibitem{lab4} A. K. Soma et al, Nucl. Instrum. Methods Phys. Res. A, {\bf 836}: 67-82 (2016); W. Zhao et al. (CDEX Collaboration), Phys. Rev. D, {\bf 88}: 052004 (2103)
\bibitem{lab5} Q. Yue et al. (CDEX Collaboration), Phys. Rev. D, {\bf90}:091701(R) (2014)
\bibitem{lab6} W. Zhao et al. (CDEX Collaboration), Phys. Rev. D, {\bf93}:092003(R) (2016)
\bibitem{lab7} K. J. Kang et al., Front. Phys. {\bf8}:412 (2013)
\bibitem{lab8} Y. C. Wu et al., J. Tsinghua Univ (Sci \& Technol), {\bf53}:1365 (2013)
\bibitem{lab9} J. L. Ma et al., Applied Radiation and Isotopes, {\bf127}:130-136 (2017)
\bibitem{lab10} L. T. Yang et al,  Nucl. Instrum. Methods Phys. Res. A, {\bf 886}: 13-23 (2018)
\bibitem{lab15} John N. Bahcall, Phys. Rev., {\bf132}:362-367 (1963)
\bibitem{lab16} Y. C. Wu et al., Chin. Phys. C, {\bf37}:086001 (2013)
\bibitem{lab17} Z. M. Zeng, \emph{Research and Application of Low Background Thermal Neutron Detection Technology}, Ph.D. thesis, Tsinghua University (2017)
\bibitem{lab177} Geant4, {\url{http://geant4.web.cern.ch/geant4/}}
\bibitem{lab18} C. Savage, G. Gelmini, P. Gondolo, and K. Freese, J. Cosmol. Astropart. Phys., {\bf04}:010 (2009)
\bibitem{lab19} J. F. Ziegler, Nucl. Instrum. Methods Phys. Res. Sect. B, {\bf219}:1027-1036 (2004); S. T. Lin et al. (TEXONO Collaboration), Phys. Rev. D, {\bf79}:061101 (2009)
\bibitem{lab20} F. Donato, N. Fornengo, and S. Scopel, Astropart. Phys., {\bf9}:247 (1998)
\bibitem{lab21} C. E. Aalseth et al. (CoGeNT Collaboration), Phys. Rev. D, {\bf88}:012002 (2013)
\bibitem{lab22} C. E. Aalseth et al., Phys. Rev. D {\bf88}:012002 (2013)
\bibitem{lab23} P. Belli et al., Phys. Rev. D, {\bf84}:055014 (2011)
\bibitem{lab24} R. Agnese et al., Phys. Rev. Lett., {\bf111}:251301 (2013)
\bibitem{lab25} E. Aprile et al. (XENON Collaboration), Phys. Rev. Lett., {\bf119}:181301 (2017)
\bibitem{lab26} D. S. Akerib et al. (LUX Collaboration), Phys. Rev. Lett. {\bf118}:021303 (2017)
\bibitem{lab266} X. Y. Cui et al. (PandaX-II Collaboration), Phys. Rev. Lett., {\bf119}:181302 (2017)
\bibitem{lab27} R. Agnese et al. (SuperCDMS Collaboration), Phys. Rev. Lett., {\bf116}:071301 (2016)
\bibitem{lab28} R. Agnese et al., Phys. Rev. Lett., {\bf112}:241302 (2014)
\bibitem{lab29} G. Angloher et al., The European Physical Journal C, {\bf76}:25 (2016)
\bibitem{lab30} J. Billard et al., Phys. Rev. D, {\bf89}:023524 (2014)
\bibitem{lab31} S. K. Liu et al. (CDEX Collaboration), Phys. Rev. D, {\bf90}:032003 (2014)
\end{thebibliography}
\end{document}